\documentstyle[aps]{revtex}

\title{ALMOST-HERMITIAN RANDOM MATRICES: APPLICATIONS TO THE THEORY OF
QUANTUM CHAOTIC SCATTERING AND BEYOND\footnote{Talk at NATO ASI Conference:  
"Supersymmetry and Trace formula: Chaos and Disorder", Cambridge 1997, to  
appear in Proceedings (J.Keating and I.Lerner, eds).
}}

\author{Yan V. Fyodorov\footnote{on leave from: Petersburg Nuclear Physics  
Institute, Gatchina 188350, Russia}}

\address{Fachbereich Physik, Universit\"{a}t-GH Essen,
D-45117 Essen, Germany}

\begin{document}
\maketitle

\begin{abstract}
It is generally accepted that statistics of energy levels in {\it closed}
chaotic quantum systems is adequately described by the theory of
Random Hermitian Matrices. Much less is known about
properties of "resonances" - generic features of open quantum systems
pertinent for understanding of scattering experiments.

In the framework of the Heidelberg approach to quantum chaotic scattering
open systems are characterized by an effective {\it non-Hermitian}
random matrix Hamiltonian $\hat{ H}_{ef}$. Complex eigenvalues of
$\hat{H}_{ef}$ are S-matrix poles (resonances).
We show how to find the mean density of these poles
(Fyodorov and Sommers) and how to use the effective Hamiltonian
to calculate autocorrelations of the photodissociation cross section
(Fyodorov and Alhassid).

In the second part of the paper we review recent results
(Fyodorov, Khoruzhenko and Sommers) on
non-Hermitian matrices with independent entries in the regime
of {\it weak Non-Hermiticity}. This regime describes a crossover from
Hermitian matrices characterized by Wigner-Dyson statistics
of real eigenvalues to strongly non-Hermitian ones whose complex
 eigenvalues were studied by Ginibre.

\end{abstract}
\section{Introduction}

As is well-known, statistics of highly excited
bound states of {\it closed} quantum chaotic systems
of quite different microscopic nature is universal. Namely, it
turns out to be independent of
the microscopic details when sampled on the energy intervals large in
comparison with the mean level separation, but smaller
than the energy scale related by the Heisenberg uncertainty principle to the
relaxation time necessary for the classically chaotic system to reach
equilibrium in the phase space \cite{AlSi}.
Moreover, the
spectral correlation functions turn out to
be exactly those which are provided by the theory of
large random matrices on the {\it local} scale determined by the typical
separation $\Delta=\langle X_i-X_{i-1}\rangle$
between neighboring eigenvalues situated
around a point $X$, with brackets standing for the statistical
averaging \cite{Bohigas}.
 Microscopic justifications of the use of random matrices
for describing the universal properties of quantum chaotic systems have been
provided recently by several groups, based both on traditional semiclassical
periodic orbit expansions \cite{per,bogomol} and on advanced
field-theoretical methods \cite{MK,aaas}.
These facts make the theory of random Hermitian matrices a powerful and versatile
tool of research in different branches of modern theoretical physics,
see e.g.\cite{Bohigas,Guhr}.

Very recently complex eigenvalues of non-Hermitian random matrices have
also attracted much research interest due to their relevance to
several branches of theoretical physics.
Most obvious motivation comes from the quantum
description of {\it open} systems \cite{Sok,FS,reso}
 whose fragments can
escape, at a given energy, to infinity or come from infinity.
For systems of this kind  the notion of
discrete energy levels loses its validity. Actually, chaotic scattering
manifests itself in terms of a high density of  poles of
the scattering matrix placed
irregularly in the complex energy plane. Each of these poles, or {\it
resonances}, $ E_{k}={\cal E}_{k}-\frac{i}{2}\Gamma_{k}$,
is characterized not only by energy
${\cal E}_{k}$  but also by a finite width $\Gamma_{k}$
defined as the imaginary part of the corresponding complex energy and
reflecting the finite lifetime of the states in the open system.
Recently, the progress in numerical techniques and
computational facilities made available high accuracy patterns of
resonance poles for realistic atomic and molecular systems in the
regime of quantum chaos, see e.g.  \cite{Main,Blumel,Man}.

Due to
the presence of these resonances, elements of the scattering matrix
 show irregular fluctuations as functions of the energy of incoming
waves, see \cite{Smilansky} and references therein.
The main goal of the theory of quantum chaotic scattering is to
provide an adequate statistical description of such a behavior.

Whereas the issue of energy level statistics
in closed chaotic systems was addressed in an
enormous amount of papers statistical characteristics of
resonances are much less studied and attracted significant
attention only recently, see \cite{Sok,FS,reso,HN}
and references in \cite{FSR}.

One possible way of doing this is to address resonances in the so-called
"Heidelberg approach" suggested in the pioneering paper \cite{VWZ}
and described in much detail in \cite{FSR}. The approach turns out to be the  
most natural framework for incorporating random matrix
description of the chaotic scattering.

The starting point of this approach is
a division of the Hilbert space of the
scattering system into two parts: the
"interaction region" and the "channel region". The channel region
is supposed to describe a situation of two fragments being apart
far enough to neglect any interaction between them. Under these
conditions their motion along the collision coordinate is
described by a superposition of incoming and outgoing plane waves
with wavevectors depending on the internal quantum states of the
fragments. We assume that at given energy $E$ exactly $M$ different
quantum states of the fragments are allowed, defining $M$
"scattering channels" numbered by the index $a$.

At the same time, the second part of the Hilbert space
is to describe the situation when fragments are close to one
another and interact strongly. Correspondingly, any wavefunction
of the system $|\Phi(E)\rangle$ can be represented as two-component vector:
$|\Phi(E)\rangle=\left(\begin{array}{c}{\bf u} \\ \psi\end{array}\right)$,
with $\bf{u}$ and $\psi$ describing the components of the
wave function inside the interaction (respectively, channel) region.

Using the standard
 methods of the scattering theory exposed in detail in \cite{FSR}
one can relate two parts of the wavefunction to one another
and finally arrive at
the following representation of the
energy-dependent scattering matrix $\hat{S}$ in terms
of an effective non-Hermitian Hamiltonian
${\cal H}_{ef}=\hat{H}-i\hat{\Gamma}$:
\begin{equation}\label{def}\begin{array}{c}
\displaystyle{S_{ab}(E)=\delta_{ab}-2i\pi\sum_{ij}
W_{ai}[E-{\cal H}_{ef}]_{ij}^{-1}W_{jb}}
\end{array}\end{equation}
with the Hermitian Hamiltonian  $\hat{H}$ describing the {\it closed}
counterpart of the open system (i.e. interaction region decoupled from the  
channel one) and the anti-Hermitian part $\hat{\Gamma}$
arising due to a coupling to open scattering channels. In this expression
 the Hamiltonian $\hat{H}$ is written in some arbitrary basis
of states $|i\rangle$, such that
$H_{ij}=\langle i|\hat{H}|j\rangle$.
 The amplitudes $W_{ai},\quad a=1,2,...,M$ are
matrix elements coupling the internal motion in an "internal"
state $|i\rangle$ to one out of M open channels $a$. One also has to
choose the anti-Hermitian part to be
$\hat{\Gamma}=\pi\sum_{a}W_{ia}W_{ja}^{*}$ in order to ensure the unitarity
of the $M\times M$ scattering matrix $\hat{S}(E)$ \cite{Mah,Kob,Liv}.

It is natural to expect, that universal properties of
open chaotic systems are inherited from the corresponding
universality of levels of their closed counterparts.
Of course, one can expect a relation of this kind only
when incoming particles stay
inside the interaction region long enough to be able
to experience the chaoticity of internal dynamics. Going from the time
domain to the energy domain, this fact suggests that
only scattering characteristics on a scale shorter than
inverse classical relaxation time on the energy shell
are expected to be universal. Another characteristic energy scale
in this domain is a typical level spacing $\Delta$ of the closed counterpart 
of our quantum open system.
Thus, we expect the scattering characteristics (in particular, the statistics
of resonances) to be universal on the scale
comparable with $\Delta$.
In contrast, smooth energy
dependence of $S-$matrix elements on a much larger energy scale must be  
system-specific.

The next step is to incorporate the random matrix
description of  quantum chaotic systems by replacing the Hamiltonian
$\hat{H}$ by a random matrix of appropriate symmetry. Namely,
chaotic systems with preserved time-reversal invariance (TRI) should be
described by matrices $H_{ij}$ which are real symmetric. Such
matrices form the Gaussian Orthogonal Ensemble, whereas for
systems with broken TRI one uses complex Hermitian matrices from
the Gaussian Unitary Ensemble \cite{Bohigas,Guhr}.

The third essential ingredient of the Heidelberg approach
is performing the ensemble averaging non-perturbatively
in the framework of the so-called
supersymmetry technique. It was invented initially by
Efetov in the context of theory of disordered
metals and the Anderson localization \cite{Efbook,my} and adjusted
for the description of quantum chaotic scattering by Verbaarschot,
Weidenm\"{u}ller and Zirnbauer \cite{VWZ}.

The Heidelberg approach turns out to be a very powerful tool for
extracting universal properties of open chaotic systems. In the first
part of the paper we outline the derivation of the mean resonance
density in the complex plane following Fyodorov and Sommers \cite{FS}.
Another recent example of the utility of
the non-Hermitian effective Hamiltonian $\hat{H}-i\hat{\Gamma}$
is that its resolvent is related to the probability for an excited
system to decay via open channels. For this reason it can be
used to calculate the statistics of such quantities as e.g. photodissociation
cross-section in the regime of quantum chaos. We present the formula
for cross-section autocorrelation function derived recently by Fyodorov
and Alhassid \cite{FA}.

  The fact that non-selfadjoint operators appear
quite  generally when one considers open systems of various types
 is known for a long time \cite{Liv}.
It is therefore not surprising that
open quantum systems were the first examples
of applications of non-Hermitian random matrices \cite{Sok,FS,FSR},
see also recent papers \cite{reso,Sav,ewa}.
Other early applications included  also
studies of dissipative quantum maps \cite{diss,Haake_book,reichl}
 and chaotic dynamics of asymmetric neural networks \cite{neural}.

Recently, however, random matrices (more generally, random linear operators)
with complex eigenvalues emerged in many other physical contexts.
Let us briefly mention the most interesting examples.
\begin{itemize}
\item The effective Hamiltonian describing a thermal motion of an isolated
vortex in disordered type-II superconductors with columnar defects
has a form of that for a quantum particle in an {\it imaginary} vector
potential ${\bf A}$:
\begin{equation}\label{hat}
\hat{H}=\frac{1}{2m}\left(-i\hbar{\bf\nabla}+i{\bf A}\right)^2+V({\bf r}),
\end{equation}
with $V({\bf r})$ being a random potential generated by defects.
The imaginary vector potential makes the Hamiltonian
to be a non-Hermitian one.
 This fact pointed out by Hatano and Nelson \cite{Nels}
gave a boost to several interesting studies \cite{nhloc,Efnonh}.
\item  A classical diffusing particle advected
by a stationary random velocity field
${\bf v}$ is described by a
non-Hermitian Fokker-Plank random operator ${\cal L}_{FP}$ \cite{pass}:
\begin{equation}
\frac{\partial}{\partial t}n({\bf r},t)=
{\cal L}_{FP}n({\bf r},t)=\left(D{\bf \nabla}^2-{\bf \nabla}
{\bf v}\right)n({\bf r},t)
\end{equation}
where $n({\bf r},t)$ is the density of particles and $D$ is the
diffusion constant.
\item
 Recent attempts to understand
the universal features of chiral symmetry breaking in Quantum Chromodynamics
required to consider quarks in a finite chemical
potential $\mu$ interacting with the Yang-Mills gauge field.
 The corresponding partition function is given by:
\begin{equation}
{\cal Z}(m,\mu)=\left\langle\prod_{f=1}^{N_f}
\mbox{det}\left(m_f+\mu\hat{\gamma}_0+\hat{\cal D}\right)
\right\rangle_{\cal A}
\end{equation}
where $\hat{\cal D}=\hat{\gamma}_{\mu}\partial_{\mu}+
i\hat{\gamma}_{\mu}{\cal A}_{\mu}$ is the Euclidean
Dirac operator in the gauge vector potential ${\cal A}$, and
$\left\langle\ldots\right\rangle_{\cal A}=\int D{\cal A}
(\ldots)\exp{-\int d^4x F_{\mu\nu}^2}$ stands for the
averaging over all configurations of the gauge field
$F_{\mu\nu}=\partial_{\mu}{\cal A}_{\nu}-\partial_{\nu}{\cal A}_{\mu}
+i[{\cal A}_{\mu},{\cal A}_{\nu}]$

A finite chemical potential $\mu$ makes the corresponding
operator to be a non-selfadjoint one with complex eigenvalues. This fact
makes a problem of numerical evaluation of the partition function by
lattice simulations to be a very difficult one. Recently it was
suggested, that some universal features of the model can be
correctly recovered if one replaces the true gauge-field averaging by
averaging over random gauge-field configurations.
As a result, one comes to a class of non-Hermitian random matrix
problems of a particular type \cite{QCD}.
\item
Finally, let us mention that
there exist several interesting links between complex eigenvalues of
non-Hermitian random matrices and systems of
interacting particles in one and two spatial dimensions \cite{int,cal}.
\end{itemize}
 At the same time, our knowledge of statistical properties of random  
non-selfadjoint matrices is quite scarce and incomplete. This fact recently
stimulated efforts of different groups to improve our understanding in this  
direction \cite{Khor}-\cite{Kus}.

Traditional mathematical treatment of random matrices
with no symmetry conditions imposed
goes back to the pioneering work by Ginibre \cite{Gin}
who determined  all the correlation functions of the
eigenvalues
in an ensemble of complex matrices with independent Gaussian entries.
The progress in the field was rather slow but steady, see
\cite{Mehta,Gir,neural,Lehm1,Forr,Edel,Oas,Bai}.

Surprisingly, all these studies
completely disregarded the existence of a nontrivial regime
of {\it weak non-Hermiticity} recognized in the work by Fyodorov,  
Khoruzhenko and Sommers \cite{FKS1},
see more detailed discussion in \cite{FKS3}.
The guiding idea to realize the existence of such a regime
comes from the experience with resonances \cite{FS}. Guided by that example
one guesses that a new regime occurs when the
imaginary part of typical eigenvalues is comparable
with a mean {\it separation} between
neighboring eigenvalues along the real axis.

One can again employ the same supersymmetry approach which
was used to study resonances and obtain the mean density
of complex eigenvalues in the regime of weak non-Hermiticity
for matrices with independent elements \cite{FKS1,FKS3}. The density turned  
out to be described by a
formula containing as two opposite limiting cases both the Wigner
semicircular density of real eigenvalues typical for Hermitian random matrices
and the uniform density of complex eigenvalues discovered
for strongly non-Hermitian random matrices already by Ginibre \cite{Gin},
in much details addressed by Girko \cite{Gir} and
studied for different cases by other authors \cite{neural,Edel,Bai}.

Very recently, Efetov \cite{Efnonh} showed the relevance of almost-Hermitian
random matrices to the very interesting problem of
 motion of flux lines in superconductors
with columnar defects \cite{Nels}. He also managed to derive
the density of complex eigenvalues for a related, but
different set of almost-symmetric real random
matrices. This development clearly shows that, apart from
being a rich and largely unexplored mathematical object,
almost-Hermitian random matrices enjoy direct physical applications
and deserve to be studied in more detail.

Actually, the non-Hermitian matrices considered in \cite{FKS1}
and [\cite{Efnonh}] are just two limiting cases of a general three-parameter
family of non-Hermitian ensembles \cite{FKS3}.
In second section of the paper we outline
 the derivation of this fact and present the resulting expression
in terms of a non-linear $\sigma$-model integral.

Although giving an important insight into the problem, the supersymmetry
non-linear $\sigma-$ model technique suffers from at least two deficiencies. 
The most essential one is that the present state
of art in the application of the supersymmetry technique gives little hope
of access to quantities describing {\it correlations}
between  different eigenvalues in the complex plane
 due to insurmountable technical difficulties.
At the same time, conventional theory of random Hermitian matrices suggests
that  these {\it universal} correlations are the most interesting features.
The second drawback is less important for a physicist, but a crucial
one for the mathematicians: at the moment
the supersymmetry technique can not be considered
as a rigorous mathematical tool and has the status of a powerful
heuristic method.

Fortunately, for the simplest case of almost-Hermitian Gaussian
random matrices one can
develop the rigorous mathematical theory based on the method of
orthogonal polynomials. Such a method is free from the above mentioned
problem and allows one to study
correlation properties of complex spectra to the same degree as
is typical for earlier studied classes of random matrices \cite{FKS2}.
We briefly discuss the obtained results in the end of the paper.
The detailed exposition of the method and the derivation of the
results can be found in \cite{FKS3}.
 Unfortunately, the paper
\cite{FKS2} contained a number of misleading misprints. For this reason we
 indicate those misprints in the present text by using footnotes.

\section{Non-Hermitian random matrices in the theory of chaotic
quantum scattering}

To calculate the density of resonance poles in the complex energy
plane we notice that they are just eigenvalues of the effective
non-Hermitian Hamiltonian introduced in Eq.(\ref{def}).

Without any loss of generality coupling amplitudes
$W_{aj}$ can be chosen in a way ensuring that the average
$S-$matrix is diagonal in the channel basis:
$\overline{S_{ab}}=\delta_{ab}\overline{S_{aa}}$.
Then one finds the following expression \cite{VWZ}:
\begin{equation}\label{sav}
\overline{S_{aa}}=\frac{1-\gamma_{a}g(X)}
{1+\gamma_{c}g(X)};\quad \gamma_{a}=\pi\sum_{i}W^{*}_{ia}W_{ia}
\end{equation}
where $g(X)=i X/2+\pi\overline{\nu(X)}$ and
$\pi\overline{\nu(X)}=(1-X^2/4)^{1/2}$ is the semicircular level
density.   The strength of coupling to
continua is convenient to characterize via the transmission
coefficients  $T_{a}=1-|\overline{S_{aa}}|^2$.
These quantities measure a part of the incoming flux in a given
channel that spends a substantial part of the time in the interaction
region \cite{VWZ,Lew}.
The case $T_a\ll 1$ corresponds to almost closed channel $a$, whereas
the opposite limiting case $T_a=1$ corresponds to the perfect
coupling between the interaction region and the channel $a$.
It is easy to
see that both limits $\gamma_{a}\to 0$ and $\gamma_{a}\to \infty$ equally
correspond to the weak effective coupling regime whereas the
strongest coupling (at fixed energy $X$ ) corresponds to $\gamma_{a}=1$.

 In the case of weak effective coupling to continua
individual resonances do not overlap: $\langle\Gamma\rangle\ll
\Delta$, with $\Delta=(\nu(X)N)^{-1}$ standing for
the mean level spacing of the
"closed" system and $\langle\Gamma\rangle$ standing
for the mean resonance width. Under these
conditions one can use a simple first order
perturbation theory to calculate resonance
widths in terms of eigenfunctions of
the closed system.
Since different components of eigenvectors
of large random matrices are decorrelated and Gaussian-distributed,
 one finds in such a procedure that
the scaled widths $y_{s}=\frac{\Gamma}{\langle\Gamma\rangle}$ are
 distributed according to the so-called $\chi^2$-distribution:
\begin{equation}\label{chi}
\rho(y_{s})=\frac{(\nu/2)^{\nu/2}}{\Gamma(\nu/2)}y_{s}^{\nu/2-1}
e^{-\frac{\nu}{2} y_{s}}
\end{equation}
 where $\Gamma(z)$ stands for the Gamma function and
the parameter $\nu=M $ $(\nu=2M)$ for systems
with preserved (broken) time reversal invariance,
$M$ being the number of open scattering channels.
The case $\nu=1$ is known as Porter-Thomas distribution and
was shown to be in agreement with experimental data
(see some references in \cite{Gasp}).

Experimentally, one quite frequently
encounter the case of only $M\sim 1$ open channels
and  $\langle \Gamma\rangle\sim \Delta$, see e.g.
\cite{Stone,Schrev,Flam}. Under this
situation resonances overlap considerably and one can
not use perturbation theory any longer.
The problem of determining the statistical characteristics of the chaotic
scattering in the regime of (partly or completely)
overlapping resonances is of essentially non-perturbative nature.
As a result, one has to use some non-perturbative
methods allowing to evaluate the ensemble averaging exactly
for arbitrary ratio
$\langle \Gamma\rangle/ \Delta$.

Fortunately, one can study very efficiently
various  universal statistical features of chaotic quantum scattering
by performing the ensemble averaging with the use of
the supersymmetry method \cite{VWZ}.

One can recover the spectral density
\begin{equation}\label{defden}
\rho(Z)=\sum_{k=1}^N\delta^{(2)}(Z-Z_k)=\sum_{k=1}^N
\delta(X-X_k)\delta(Y-Y_k)\equiv \rho(X,Y)
\end{equation}
of complex eigenvalues $Z_k=X_k+iY_k, \quad k=1,2,...,N$
 if one knows the "potential" \cite{neural}:
$$
\Phi(X,Y,\kappa)=\overline{\frac{1}{2\pi}
\ln{Det[(Z-{\cal H}_{ef})(Z-{\cal H}_{ef})^{\dagger}+\kappa^2]}}
$$
in view of the relation:$\rho(X,Y)=
\lim_{\kappa\to 0}\partial^2 \,\Phi(X,Y,\kappa)$,
where $\partial^2$ stands for the two-dimensional Laplacian.
Technically, it is convenient to introduce
 the generating function (cf.\cite{FSR})
\begin{equation}\label{genf}
{\Large \cal Z}=
\frac{{\mbox Det}\left[(Z-{\cal H}_{ef})(Z-{\cal  
H}_{ef})^{\dagger}+\kappa^2\right]}
{{\mbox Det}\left[(Z_b-{\cal H}_{ef})(Z_b-{\cal  
H}_{ef})^{\dagger}+\kappa^2\right]}
\end{equation}
in terms of which
\begin{equation}
\rho(Z)=-\frac{1}{\pi}\lim_{\kappa\to 0}\frac{\partial}{\partial Z^*}
\lim_{Z_b\to Z}\frac{\partial}{\partial Z_b}{\Large\cal Z}
\end{equation}
To facilitate the ensemble averaging
 we follow the standard route and represent the ratio of the two
determinants in Eq.(\ref{genf}) in terms of a Gaussian
superintegral
over eight-component supervectors
$\Phi_{i}=\left(\begin{array}{c}\Psi_{i}(+)\\ \Psi_{i}(-)\end{array}\right)$
where
$ \Psi_{i}(\pm)=\left(\begin{array}{c} \vec{R}_{i}(\pm)\\
\vec{\eta}_{i}(\pm)\end{array}
\right)$ and
$$ \vec{R}_{i}(\pm)=\left(\begin{array}{c}
r_{i}(\pm)\\r_{i}^{*}(\pm)\end{array}\right)\quad;\quad
\vec{\eta}_{i}(\pm)=
\left(\begin{array}{c}\chi_{i}(\pm)\\ \chi_{i}^{*}(\pm)\end{array}\right)
\quad;\quad
\vec{\eta}_{i}^{\dagger}(\pm)=
\left(\chi_{i}^{*}(\pm);- \chi_{i}(\pm)\right)$$
with components $r_{i}(+),r_{i}(-);\quad i=1,2,...,N$ being complex commuting
variables and $\chi_{i}(+),\chi_{i}(-)$ forming the corresponding Grassmann
parts of the supervectors $\Psi_{i}(\pm)$.

 Further evaluation goes along the lines discussed in
\cite{FSR} in more detail.
After a set of standard manipulations one
arrives at the following expression for the density $\rho_X(y)
\equiv\frac{1}{\pi}\rho(X,Y)\Delta^2(X)$ of scaled
resonance widths $y=\frac{\pi\Gamma}{\Delta}$ (measured in units of the local 
mean level spacing $\Delta(X)$ of the closed system) for
the resonances whose positions are within
a narrow window around the point $X$ of the spectrum:
\begin{eqnarray}\label{widres}
&&\langle\rho_X(y)\rangle=\frac{1}{16}\int d\mu(\hat{Q})
\mbox{Str}\left(\hat{\sigma}_\tau^{(F)}\hat{Q}\right)
\mbox{Str}\left(\hat{\sigma}_\tau\hat{Q}\right)
\exp{\frac{i}{4}y\mbox{Str}\left(\hat{\sigma}_\tau\hat{Q}\right)}
\prod_{a=1}^M\mbox{ Sdet}^{-1/4}\left[1-\frac{i}{2g_a}
\left\{\hat{Q},\hat{\sigma}_{\tau}\right\}\right]
\end{eqnarray}
Here the integration goes over the set of $8\times 8$
supermatrices $\hat{Q}$ satisfying the constraint $\hat{Q}^2=-1$
and $\left\{\hat{Q},\hat{\sigma}_{\tau}\right\}=
\hat{Q}\hat{\sigma}_\tau+\hat{\sigma}_\tau\hat{Q}$ stands
for the anticommutator.
Properties of these matrices and the integration measure
$d\mu(\hat{Q})$ can be found in \cite{Efbook}.
Other $8\times 8$ supermatrices entering the expression
Eq.(\ref{widres}) are as follows:
$$
\hat{\sigma}_{\tau}^{(F)}=
\left(\begin{array}{cc}\hat{0}_4&\hat{\tau}_3^{(F)}\\
\hat{\tau}_3^{(F)}&\hat{0}_4\end{array}\right);
\quad \hat{\sigma}_{\tau}=\left(\begin{array}{cc}0&\hat{\tau}_3\\
\hat{\tau}_3&0\end{array}\right)
$$
and $\hat{\tau}_3;\,,\hat{\tau}_3^{(F)}$ are $4\times 4$
diagonal supermatrices: $\hat{\tau}_3=\mbox{diag}\{\hat{\tau},\hat{\tau}\};
\,\,\hat{\tau}_3^{(F)}=\mbox{diag}\{\hat{0}_2,\hat{\tau}\}$.
with $ \hat{\tau}=\mbox{diag}(1,-1)$.
We also introduced quantities
$g_a=\frac{1}{2\pi\overline{\rho}(X)}
(\gamma_{a}+\gamma_{a}^{-1})$ related to the transmission coefficients as  
$g_a=2/T_a-1$
 and used the symbols $\mbox{Str,\,Sdet}$ for the graded
trace and the graded determinant, correspondingly.

The expression above is valid for chaotic systems with preserved as
well as with broken time-reversal invariance. To extract the explicit form
of the distribution function one still has to perform the integration
over the manifold of the supermatrices $Q$ which is different for two cases. 
In general it is a rather
difficult calculation due to a cumbersome parametrisation of that
manifold. At the moment the result is known for the simplest case
of the systems with broken time-reversal invariance \cite{FS,FSR}.
For the sake of simplicity we present this distribution for the case of  
equivalent channels $a=1,...,M$ with equal transmission coefficients $T_a=T$.

First of all, it turns out that the
mean resonance width is related to the transmission coefficient $T$
as:
\begin{equation}\label{Simonius}
\langle\Gamma\rangle=-\Delta\frac{M}{2\pi}\ln{(1-T)}\equiv
\Delta\frac{M}{2\pi}\ln{\frac{g+1}{g-1}}
\end{equation}

The formula Eq.(\ref{Simonius}) is well known in nuclear physics as   
Moldauer-Simonius relation \cite{Mol}.

It is convenient to use the parameter $\kappa=-\frac{M}{2}\ln{(1-T)}$ as a  
measure of the resonance overlap. It is related to the mean widths  as
$\kappa = \pi\frac{\langle \Gamma\rangle}{\Delta}$  and therefore gives a  
typical number of neighboring resonances that overlap substantially.

Measuring the resonance widths in units of the mean widths $\langle\Gamma\rangle$
one finds the following distribution function.

\begin{eqnarray}\label{main}
\rho\left(y_s=\frac{\Gamma}{\langle \Gamma\rangle}\right)&=&
\frac{1}{2\Gamma(M)\kappa y^2}\int_{y_sb_{-}}^{y_sb_{+}} dt t^M e^{-t}\\
&=& \nonumber(-1)^M\frac{y_s^{M-1}}{\Gamma(M)}\frac{d^M}{dy_s^M}
\left(\exp{-[\kappa\coth{(\kappa/M)}y_s]}
\frac{\sinh{\kappa y_s}}{\kappa y_s}\right).
\end{eqnarray}
where we used the notations $b_{\pm}=\kappa e^{\pm  
\kappa/M}/\sinh{(\kappa/M)}$ and $\Gamma(M)=(M-1)!$ for the Euler  
gamma-function..

Properties of this distribution are discussed in much details in
\cite{FSR}, also for the case of non-equivalent channels.
Let us only briefly mention the most interesting features.

Quick inspection of eq.(\ref{main}) shows that it is indeed
reduced to the $\chi^2$
distribution, eq.(\ref{chi}) when the effective
coupling to continua is weak: $\kappa\ll 1$ . Under this condition
resonances are typically too narrow to overlap with others:
$\Gamma\ll \Delta$.
However, as long as the effective coupling
becomes stronger: $T\to 1$, hence $g\gg 1$ the parameter $\kappa$ grows large.
Under these conditions another domain of resonance widths
becomes more and more important:
$$e^{-\kappa/M}\frac{\sinh{(\kappa/M)}}{(\kappa/M)}< y_s <  
e^{\kappa/M}\frac{\sinh{(\kappa/M)}}{(\kappa/M)},$$ where the distribution
eq.(\ref{main}) shows the powerlaw decrease:
$\rho(y_s)\approx \frac{1}{2\kappa}My_s^{-2}$.
The most drastic difference from eq.(\ref{chi}) occurs for
the maximal effective coupling
$g=1$ (i.e $\kappa=\infty$). In this regime the powerlaw tail
extends up to infinity, making all positive moments
( starting from the first one) to be apparently divergent.
 One can argue that the powerlaw tail $My_s^{-2}$ turns out to be
dictated by classical processes of exponential escape typical for
fully chaotic systems \cite{FSR}. The rate of this escape in the  
semiclassical limit $M\gg 1$ is provided by the value of the gap in the  
distribution of
resonance width, see \cite{Sok} for a more detailed discussion.

The best candidates for checking the applicability of eq.(\ref{main})
to real physical systems are realistic models of ballistic mesoscopic devices
subject to an applied magnetic field that serves to break the TRI \cite{Stone}.
It is however quite clear that all the basic qualitative features of the
distribution eq.(\ref{main}) (in particular, the powerlaw
behavior $\rho(y_s)\propto My_s^{-2}$ for the overlapping
resonance regime) should
be valid for the systems with preserved TRI as well. Recent numerical data
\cite{Seba} support the validity of this conjecture.

We have seen, that the  non-Hermitian random matrix Hamitonian
$\hat{H}_{ef}$ appeared naturally in the scattering matrix
description of open quantum systems.
Actually, such a Hamiltonian is the most adequate tool to
describe the quantum relaxation processes such as escape of
the particle from the interaction region in the regime of quantum chaos.
 Some aspects of such a relaxation were studied some
time ago in \cite{Harney} and reconsidered in more details
recently by Savin and Sokolov \cite{Sav} who used insights
provided by the resonance widths distribution Eq.(\ref{main})

It is therefore quite natural that the resolvent of
the non-Hermitian effective Hamiltonian $\hat{H}_{ef}=\hat{H}-i\hat{\Gamma}$
 is related to the probability for an excited
system to decay via one of open channels. For this reason it can be
used to calculate the statistics of such quantities as the Wigner {\it
time delay} \cite{td1} which is a measure of mean time spent by a scattered
particle inside the interaction region.

One more example of the utility of the resolvent of the non-Hermitian
effective Hamiltonian $\hat{H}_{ef}$ is that it can be related to
the total photodissociation
crossection in the regime of quantum chaos. The formula
for the cross section autocorrelation function was derived recently by Fyodorov
and Alhassid \cite{FA}. Below we outline the derivation and present
the final result.

The total energy-dependent cross section $\sigma(E)$
is defined as a probability to be excited from a ground state $|g\rangle$ and to
dissociate at a given energy $E$ per unit time and
per unit incoming photon flux density.
In the dipole approximation it is given by (see, e.g.
detailed discussion in \cite{Schinke}) :
\begin{equation}\label{cross}
\sigma(E)=\sigma_0\sum_a\left|\left\langle g|{\bf
\hat{\mu}}|\Phi_a^{out}(E)\right\rangle\right|^2
\end{equation}
Here ${\bf \hat{\mu}}$ is the dipole operator ${\bf \hat{\mu}}=e{\bf \cal E}
{\bf r}$ of the system in the external electric field
${\bf\cal E}$, $\sigma_0$ is a constant proportional to the excitation
energy $E$ and $\left|\Phi_a^{out}(E)\right\rangle$ is the exact wave
function of the system at energy $E$ subject to {\it outgoing}
boundary conditions in one of the open  channels $a=1,2,...,M$.

It turns out, that for many systems of interest
(e.g. molecules $HO_2$\cite{Schrev} and $H_3^{+}$\cite{Man}),
the cross section patterns look like
irregular fluctuating signals consisting of
many randomly positioned (partly) overlapping resonance peaks.
Such a behavior ( typical also for the resonance scattering
in atomic systems \cite{Main,Flam}) has its origin in
the underlying pattern of resonances in the complex plane.

We already mentioned that one can relate the "internal" and "external"
parts of the wavefunctions by using the resolvent of the non-Hermitian  
effective Hamiltonian
$\hat{H}_{ef}$. In the present case such a relation can be writen as (see  
e.g. \cite{Sok1}):
\begin{equation}\label{int}
{\bf u}=\left(E-\hat{H}_{ef}^{\dagger}\right)^{-1}\hat{W}{\bf
B}
\end{equation}
where $M$-component vector ${\bf B}$ contains amplitudes of outgoing
waves in each of the open channels.

The ground state wavefunction
describes a bound state and as such has no components outside the
interaction region.
Using this fact and Eq.(\ref{int}) one finds after
some algebraic manipulations  that the cross section Eq.(\ref{cross})
can be rewritten in the following convenient form:
\begin{equation}\label{im}
\sigma(E)\propto Im \left\langle g\left|{\bf \mu}\frac{1}{E-{\cal H}_{ef}}{\bf
\mu} \right|g\right\rangle
\end{equation}
which is just a particular case of the optical theorem.
One also can arrive at the expression Eq.(\ref{im})
by resumming the perturbation theory, see \cite{Flam} for
more details and relevant references.

The advantage of the form Eq.(\ref{im}) is that it expresses the
photodissociation cross section in terms of the resolvent of an
effective non-Hermitian operator ${\cal
H}_{ef}=H_{in}-i\pi WW^{\dagger}$ which is known to describe open
chaotic systems in the random matrix formalism. It allows to apply very well  
developed methods of evaluating averages of products of
resolvents based on the Efetov supermatrix formalism.
 Measuring energy separations in units of the mean level spacing
of the closed system $\Delta$
 one finds in such a calculation the cross section
autocorelation function \cite{FA}:
$$
S\left(\omega=2\pi\Omega/\Delta\right)=
\frac{\langle\sigma(E-\Omega/2)\sigma(E+\Omega/2)\rangle}
{\langle\sigma(E)\rangle^2}-1,
$$
 to be a sum of two terms
$S(\omega)=S_{1}(\omega)+S_{2}(\omega)$
given by the following expressions:
\begin{equation}\label{main1}\begin{array}{c}
\displaystyle{S_{1,2}(\omega)=
\int_{-1}^1d\lambda\int_{1}^{\infty}d\lambda_1\int_{1}^{\infty}d\lambda_2
\frac{\cos{[\omega(\lambda_1\lambda_2-\lambda)]}(1-\lambda^2)}
{[\lambda_1^2+\lambda_2^2+\lambda^2-2\lambda_1\lambda_2\lambda-1]^2}}

\displaystyle{f_{1,2}(\lambda,\lambda_1,\lambda_2) \prod_{c=1}^{M}
\frac{(g_a+\lambda)}
{[(g_a+\lambda_1\lambda_2)^2-(\lambda_1^2-1)(\lambda_2^2-1)]^{1/2}}}
\end{array}
\end{equation}
where
$$
f_1(\lambda,\lambda_1,\lambda_2)=(\lambda_1\lambda_2-\lambda)^2;
\quad
f_2(\lambda,\lambda_1,\lambda_2)=2\lambda_1^2\lambda_2^2-\lambda_1^2
-\lambda_2^2-\lambda^2+1
$$
The parameters $g_a$ were introduced before and
related to the transmission coefficients as
$g_a=2/T_a-1$.

It is worth mentioning that each of the contributions $S(\omega)_{1,2}$
represent an interesting object by itself.
Namely, $S_1(\omega)$ coincides with the autocorrelation function
of the so-called Wigner time delays studied in some detail in \cite{td1},
whereas $S_2(\omega)$ is related by the Fourier-transform to
the so-called "norm leakage" out the interaction region.
The latter quantity was introduced recently by
Savin and Sokolov as a characteristic
of the process of quantum relaxation in chaotic systems
and studied in detail for the simplest case of broken
time-reversal invariance \cite{Sav}.

Actually, the starting Fermi golden rule formula Eq.(\ref{cross})
is valid for an arbitrary excitation of the system with a weak perturbation
$\hat{\mu}$.
For this reason the autocorrelation function of crossections
presented above is also of a general applicability.

Finally, it is necessary to mention that in the limit $T_a=0$ for all $a$
(corresponding to a closed system with purely
bound spectra and no possibility for photodissociation)
the expression Eq.(\ref{main1})
reduces to the "oscillator strength" correlation function
 found by Taniguchi et al. \cite{Tan}.

\section{Non-Hermitian matrices with independent elements:
Universal properties in the regime of weak non-Hermiticity}

To begin with, any $N\times N$ matrix $\hat{J}$ can be decomposed
into a sum of its Hermitian and skew-Hermitian parts:
$
\hat{J}=\hat{H}_1+i\hat{H}_2,
$
where  $\hat{H}_1=(\hat{J}+\hat{J}^\dagger )/2 $ and $\hat{H}_2
=(\hat{J}-\hat{J}^\dagger )/2i$. Following this, we
consinder an ensemble of random $N\times N$ complex matrices
$\hat{J}=\hat{H}_1+iv\hat{H}_2$
where $\hat{H}_p;\,\, p=1,2$ are both Hermitian:  
$\hat{H}^{\dagger}_p=\hat{H}_p$. The parameter $v$ is used to control the  
degree of non-Hermiticity.

In turn, complex Hermitian matrices $\hat{H}_p$
can always be represented as $\hat{H}_1=\hat{S}_1+iu\hat{A}_1$ and
$\hat{H}_2=\hat{S}_2+iw\hat{A}_2$, where $\hat{S}_p=\hat{S}_p^T$
is a real symmetric matrix, and $\hat{A}_p=-\hat{A}_p^T$
is a real antisymmetric one. From this point of view the parameters
$u,w$ control the degree of being non-symmetric.

Throughout the paper we consider the matrices $\hat{S}_1,\hat{S}_2,\hat{A}_1,
\hat{A}_2$ to be mutually statistically independent, with i.i.d. entries
normalized in such a way that:
\begin{equation}\label{norm}
\lim_{N\to\infty}\frac{1}{N}{\mbox  
Tr}\hat{S}_p^2=\lim_{N\to\infty}\frac{1}{N}{\mbox Tr}\hat{A}_p\hat{A}_p^T=1
\end{equation}

As is well-known \cite{Bohigas}, this normalisation ensures
that for any value of the parameter $u\ne 0$ , such that $u=O(1)$
when $N\to \infty$ statistics
of real eigenvalues of the Hermitian matrix of the form
$\hat{H}=\hat{S}+iu\hat{A}$ is identical (up to a trivial rescaling) to
that of $u=1$, the latter case known as the Gaussian Unitary Ensemble (GUE).  
On the other hand, for $u\equiv 0$ real eigenvalues of the real symmetric  
matrix $\hat{S}$
follow another pattern of the so-called Gaussian Orthogonal Ensemble (GOE).

The non-trivial crossover between GUE and GOE types of statistical
behavior happens on a scale $u\propto 1/N^{1/2}$ \cite{cross}.
This scaling can be easily
understood by purely perturbative arguments \cite{Alt}.
Namely, for $u\propto 1/N^{1/2}$ a typical shift
$\delta \lambda$ of eigenvalues of the symmetric matrix $S$ due to
the antisymmetric perturbation  $iu\hat{A}$ is of the same
order as the mean spacing $\Delta $ between unperturbed eigenvalues
: $\delta\lambda\sim \Delta\sim 1/N$.

Similar perturbative arguments show \cite{FKS1}, that the
most interesting behavior
of {\it complex} eigenvalues of non-Hermitian matrices
should be expected for the parameter $v$ being scaled in
a similar way: $v\propto 1/N^{1/2}$.
It is just the regime when the {\it imaginary} part
${\mbox Im} Z_k $ of a typical eigenvalue $Z_k$ due to
non-Hermitian perturbation is of the same order as the
mean spacing $\Delta $ between unperturbed real eigenvalues :
${\mbox Im}Z_k \sim \Delta\sim 1/N$.
Under these conditions a non-Hermitian matrix $J$ still
"remembers" the statistics of its Hermitian part $\hat{H}_1$.
As will be clear afterwards, the parameter $w$
should be kept of the order of unity in order to influence the statistics
of the complex eigenvalues.

It is just this regime of {\it weak non-Hermiticity} which we are interested in.
Correspondingly, we scale the parameters as
\footnote{In the Letter \cite{FKS2} there is a misprint in the
definition of the parameter $\alpha$.}:
\begin{equation}\label{scale}
v=\frac{\alpha}{2\sqrt{N}};\quad u=\frac{\phi}{2\sqrt{N}}
\end{equation}
and consider $\alpha,\phi,w$ fixed of the order O(1) when $N\to \infty$.

To be specific, we consider the real symmetric matrix $\hat{S}_1$ to be  
taken from the ensemble of {\it sparse} random matrices \cite{MF,FSC}
characterized by the following probability density of a given entry $S_{ij}$ :
\begin{equation}\label{distr}
{\cal P}(S_{ij})=(1-\frac{p}{N})\delta(S_{ij})+\frac{p}{N}h(S_{ij})
\end{equation}
where $h(s)=h(-s)$ is an arbitrary even distribution function satisfying the
conditions:
$h(0)<\infty;\quad \int h(s) s^{2} ds<\infty$ and $p$ stands for the mean value
of non-zero matrix elements per column.
Actually, this ensemble is the most general one among those with independent 
elements, and statistics of its eigenvalues was proved to be completely  
universal \cite{MF,note}, up to a rescaling by {\it ensemble-dependent}
mean eigenvalue density $\nu(X)$. Statistics of the matrix elements of all  
other matrices
$S_2,A_{1,2}$ is immaterial as long as their elements are statistically  
independent as well.

The calculation of the mean density of complex eigenvalues follows essentially
the same route as that outlined in the previous section for the resonances.
The method used \cite{FKS3}
 is a generalization of the Efetov's technique to the case
of sparse random matrices suggested in \cite{MF}
 (see some details also in \cite{FSC}).
As the result, one arrives at the following expression \cite{FKS3}:

\begin{equation}\label{unires}
\langle\rho(X,y)\rangle=\frac{N\nu(X)}{16}
\int d\mu(\hat{Q})\mbox{Str}\left(\hat{\sigma}_\tau^{(F)}\hat{Q}\right)
\mbox{Str}\left(\hat{\sigma}_\tau\hat{Q}\right)\exp{-S(\hat{Q})}
\end{equation}
\begin{equation}
S(\hat{Q})=-\frac{i}{2}y\mbox{Str}\left(\hat{\sigma}_\tau\hat{Q}\right)-
\frac{a^2}{16}\mbox{Str}\left(\hat{\sigma}_\tau\hat{Q}\right)^2+
\frac{b^2}{16}\mbox{Str}\left(\hat{\tau}_2\hat{Q}\right)^2-\frac{c^2}{16}
\mbox{Str}\left(\hat{\sigma}\hat{Q}\right)^2
\end{equation}
where we introduced the scaled imaginary parts $y=\pi\nu(X)NY$
and used the notations: $a^2=\left(\pi\nu(X)\alpha\right)^2,
\quad b^2=\left(\pi\nu(X)\phi\right)^2,
\quad c^2=\left(\pi\nu(X)\alpha w\right)^2$.
The supermatrices $\hat{\tau}_2$ and $\hat{\sigma}$
 entering this expressions are as follows:
$$
\hat{\tau}_2=\mbox{diag}\{\hat{\tau}_3,\hat{\tau}_3\};\quad  
\sigma=\left(\begin{array}{cc}0&\hat{I}_4\\ \hat{I}_4&0\end{array}\right)
$$
and the supermatrices $\hat{\sigma}_{\tau}$ and $\hat{\tau}_3$ were defined
after Eq.(\ref{widres}).
The expression (\ref{unires}) is just the universal $\sigma-$ model
representation of the mean density of complex eigenvalues in the
regime of weak non-Hermiticity we were looking for.
The universality is clearly manifest: all the particular details about the
ensembles entered only in the form of mean density of
real eigenvalues $\nu(X)$.  The density of complex
eigenvalues turns out to be dependent on three parameters:
$a,b$ and $c$, controlling the degree of non-Hermiticity
($a$), and symmetry properties of the
Hermitian part ($b$) and non-Hermitian part ($c$).

The following comment is appropriate here. The derivation above
can be done not only for ensembles with i.i.d. entries but also for any  
"rotationaly invariant" ensemble of real symmetric matrices $\hat{S}_1$.
To do so one can employ the procedure invented by Hackenbroich and
Weidenm\"{u}ller \cite{HW} allowing one to map the
correlation functions of the invariant
ensembles (plus perturbations) to that of Efetov's $\sigma-$model.

Still, in order to get an explicit expression for the
density of complex eigenvalues one has to
evaluate the integral over the set of supermatrices
$\hat{Q}$. In general, it is an elaborate
task due to complexity of that manifold.

At the present moment such an evaluation was successfully performed for two  
important cases: those of almost-Hermitian matrices and real almost-symmetric  

matrices. The first case ( which is technically the simplest one)
corresponds to $\phi\to\infty$, that is $b\to \infty$.
Under this condition only that part of the matrix
$\hat{Q}$ which commutes with
$\hat{\tau}_2$ provides a nonvanishing contribution. As the result,
$\mbox{Str}\left(\hat{\sigma}\hat{Q}\right)^2=\mbox{Str}\left(\hat{\sigma}_{\tau}\hat{Q}\right)^2$  
so that second and fourth term in Eq.(\ref{unires})
can be combined together. Evaluating the resulting integral,
and introducing the notation $\tilde{a}^2=a^2+c^2$ one finds \cite{FKS1}:
\begin{eqnarray}\label{13}
\rho_X(y)=\sqrt{\frac{2}{\pi}}\frac{1}{\tilde{a}}\exp
 \left( -\frac{2y^2}{\tilde{a}^2} \right)
\int_0^1  dt \cosh (2ty) \exp{(-\tilde{a}^2t^2/2)},
\end{eqnarray}
where $\rho_X(y)$ is the density of the scaled imaginary parts $y$
for those eigenvalues, whose real parts are situated around the point
$X$ of the spectrum (cf. Eq.(\ref{widres})).

It is easy to see, that when $\tilde{a}$ is large one can effectively
put the upper boundary of integration in Eq.(\ref{13}) to be infinity
due to the Gaussian cut-off of the integrand. This immediately
results in the uniform density $\rho_X(y)=(\tilde{a}^2)^{-1}$ inside the
interval $|y|<\tilde{a}^2/2$ and zero otherwise. Translating this result to the
two-dimensional density of the original variables $X,Y$, we get:
\begin{equation}\label{girko}
\rho(X,Y)=\left\{\begin{array}{cc} \frac{N}{4\pi v^2(1+w^2)}& \mbox{for}
\,\,|Y|\le 2\pi\nu(X)v^2(1+w^2)\\ 0&\mbox{otherwise}\end{array}\right.
\end{equation}

This result is a natural generalization of the so-called "ellipic law" known
for strongly non-Hermitian random matrices \cite{Gin,Gir,neural}.
Indeed, the curve encircling the domain of the uniform eigenvalue
density is an ellipse: $\frac{Y^2}{2v^2(1+w^2)}+\frac{X^2}{4}=1$ as long as the
mean eigenvalue density of the Hermitian counterpart is given by the
semicircular law. The semicircular density is known to be
shared by ensembles with i.i.d. entries, provided the mean
number $p$ of
non-zero elements per row grows with the matrix size as $p\propto
N^{\alpha}; \,\, \alpha>0$, see \cite{MF}. In the general case of
sparse or "rotationally invariant" ensembles the function $\nu(X)$
might be quite different from the semicircular law.
Under these conditions Eq.(\ref{girko})
still provides us with the corresponding density of complex eigenvalues.

The second nontrivial case for
which the result is known explicitly is due to Efetov \cite{Efnonh}.
It is the limit
of slightly asymmetric real matrices corresponding in the present notations to:
$\phi\to 0; w\to \infty$ in such a way that the product
$\phi w=\tilde{c}$ is kept fixed.
The density of complex eigenvalues turns out to be given by:
\begin{eqnarray}\label{Efres}
\rho_X(y)=\delta(y)\int_0^1 dt\exp{(-\tilde{c}^2t^2/2)}+
2\sqrt{\frac{2} {\pi}}\frac{|y|}{\tilde{c}}
\int_1^{\infty}du \exp \left(-\frac{2y^2u^2}{\tilde{c}^2} \right)
\int_0^1  dt t \sinh (2t|y|) \exp{(-\tilde{c}^2t^2/2)},
\end{eqnarray}

The first term in this expression shows that everywhere in the regime of
"weak asymmetry" $\tilde{c}<\infty$ a finite fraction of
eigenvalues remains on the real axis.

Such a behavior is qualitatively
different from that typical for the
case of "weak non-Hermiticity" $\tilde{a}<\infty$, where eigenvalues
acquire a nonzero imaginary part with probability one.

In the limit $\tilde{c}>>1$ the portion of real eigenvalues
behaves like $\tilde{c}^{-1}$. Remembering the normalization
of the parameter $v$, Eq.(\ref{norm}), it is easy to see that
for the case of $v=O(1)$ the number of real eigenvalues should scale
as $\sqrt{N}$.
 The fact that of the order of $N^{1/2}$ eigenvalues of strongly asymmetric
real matrices stays real was first found numerically by
Sommers et al. \cite{neural,Lehm1}, and proved by Edelman \cite{Edel}.

\section{Gaussian almost-Hermitian matrices: from Wigner-Dyson to
Ginibre eigenvalue statistics.}
In the present section we concentrate on the particular
case of almost-Hermitian random matrices with i.i.d. entries
$\hat{J}=\hat{H}_1+iv\hat{H}_2,$
 where $\hat H_1 $ and
$\hat H_2$ are taken independently from the
Gaussian Unitary Ensemble (GUE) of {\it Hermitian } matrices
with the probability density
${\cal P}(\hat{X})=Q_N^{-1}
\exp{\Big(-N/2J_0^{2}\ \mbox{Tr}\ \hat{X}^2 \Big)}$,
$\hat X = \hat X^\dagger$.

Let us now introduce a new parameter $\tau=(1-v^2)/(1+v^2)$
and choose the scale constant $J_0^2$ to be equal to
$(1+\tau)/2$, for the sake of convenience.
The parameter $\tau$ controls the magnitude of
correlation between
$J_{jk}$ and $J_{kj}$: $\langle J_{jk}J_{kj} \rangle = \tau /N $,
hence
the degree of non-Hermiticity.
This is easily seen from the probability density function
for our ensemble of the random matrices $\hat{J}$:
\begin{eqnarray}
{\cal P}(\hat{J})
=C_N^{-1} \exp \Big[-\frac{N}{(1-\tau^2)} \ \mbox{Tr}
(\hat{J}\hat{J}^\dagger -\tau \ \mbox{Re}\  \hat{J}^2) \Big],
\label{P(J)}
\end{eqnarray}
where $C_N=[\pi^2(1-\tau^2)/N^2]^{N^2/2}$.
All the $J_{jk}$  have zero mean and variance
$\langle |J_{jk}|^2 \rangle =1/N$ and only
$J_{jk}$ and $J_{kj}$ are pairwise correlated.
If $\tau =0$ all the $J_{jk}$
are mutually independent and
we have maximum non-Hermiticity.
When $\tau $ approaches unity, $J_{jk}$ and
$J_{kj}^*$ are related via $J_{jk}=J_{kj}^*$
and we are back to an ensemble of Hermitian matrices.

Our first goal is to determine the $n$-eigenvalue correlation functions
in the ensemble of random matrices specified by Eq.\ (\ref{P(J)}).
The density of the joint distribution of
eigenvalues in the ensemble is given by
\begin{eqnarray}\label{P(Z)}
{\cal P}_N(Z_1, \ldots, Z_N)= \frac{N^{N(N+1)/2}}{\pi^N 1!
\cdots N! (1-\tau^2)^{N/2}}
\exp \Big\{\frac{-N}{1-\tau^2} \sum_{j=1}^N
\Big[|Z_j|^2 - \frac{\tau}{2}(Z_j^2+{Z_j^*}^2) \Big]  \Big\} \
\prod_{j<k}|Z_j-Z_k|^2 .
\end{eqnarray}

To derive Eq.~(\ref{P(Z)}) we integrate
${\cal P}(\hat{J})$ from Eq.~(\ref{P(J)}) over the surface of all complex matrices
whose eigenvalues are $Z_1, \ldots Z_N$.
Following Dyson \cite{Mehta},
p.501, see also \cite{Edel}) we decompose
every complex matrix with distinct eigenvalues as
$ \hat{J}=\hat{U}(\hat{Z}+\hat{R})\hat{U}^{\dagger}$,
where $\hat{Z}=\mbox{diag}\{ Z_1, \ldots Z_N \}$,
$\hat {U}$ is  a unitary matrix, and
$\hat{R}$ is a strictly upper-triangular one.
If we label the eigenvalues and
require the first
non-zero element in each column of $\hat{U}$ to be positive, then
the decomposition is unique.
The Jacobian of the transformation $\hat{J}\to \{\hat{Z},\hat{R} ,\hat{U}\}$  
depends only on $\hat{Z}$ and is given by
the squared modulus of the Vandermonde determinant. So, integrating out
$\hat R$ and $\hat U$ is straightforward and the resulting expression is
Eq.~(\ref{P(Z)}).

The form of the distribution
Eq.~(\ref{P(Z)}) allows one to employ
the powerful method of orthogonal polynomials \cite{Mehta}.
Let $H_n(z)$ denote the
$n$-th Hermite polynomial,
\begin{eqnarray} \label{H}
H_n(z)=\frac{(\pm i)^n}
{\sqrt{2\pi}}\! \exp{\left(\! \frac{z^2}{2}\! \right)}
\int_{-\infty}^{\infty}\!
dt\  t^n \exp{\left(-\frac{t^2}{2}\mp izt\right)}.
\end{eqnarray}
The
crucial observation borrowed from the paper \cite{orth}
(see also the related paper \cite{FJ})
 is that the polynomials
\begin{eqnarray}\label{p_n}
p_n(Z)=\frac{\tau^{n/2} \sqrt{N} }
{\sqrt{\pi }\sqrt{ n!}(1-\tau^2 )^{1/4}}
H_n\left( \sqrt{ \frac{N}{\tau}}Z\right),
\end{eqnarray}
$n=0,1,2, \ldots $,
are orthogonal in the {\it complex plane}
$Z=X+iY$
with the weight function
\[
w^2(Z)=\exp{\left\{-\frac{ N}{(1-\tau^2)}\left[|Z|^2 -
\frac{\tau}{2}(Z^2+{Z^*}^2) \right]\right\}},
\]
i.~e. $
\int d^2Z p_n(Z)p_m(Z^*)w^2(Z)  = \delta_{nm}$, where
$d^2 Z = dX dY$.
The standard machinery of the method of orthogonal polynomials \cite{Mehta}
yields the $n$-eigenvalue correlation functions
\begin{eqnarray}
R_n(Z_1,...,Z_n)=\frac{N!}{(N-n)!}\int d^2Z_{n+1}...d^2Z_N
{\cal P}_N\{Z\}
\label{R_n}
\end{eqnarray}
in the form
\begin{eqnarray*}
R_n(Z_1,...,Z_n)&=&\det \left[ K_N(Z_j,Z_k^*)\right]_{j,k=1}^n,
\end{eqnarray*}
where the kernel $K_N(Z_1,Z_2^*)$ is given by
\begin{eqnarray}
K_N(Z_1,Z_2^*)&=w(Z_1)w(Z_2^*)&\sum_{n=0}^{N-1}p_n(Z_1)p_n(Z_2^*).
\label{K}
\end{eqnarray}

With Eqs.\ (\ref{p_n})--(\ref{K}) in hand, let us first examine
the regime of strong non-Hermiticity, i.e. the case when
$\lim_{N\to \infty} (1-\tau ) > 0$. In this regime
the averaged density of eigenvalues $N^{-1}R_1(Z)$ is asymptotically
zero outside the ellipse $[\mbox{Re} z/(1+\tau )]^2
+[\mbox{Im} z/(1-\tau )]^2 \le 1$. Inside the ellipse
$\lim_{N\to \infty} N^{-1}R_1(Z) = [\pi (1-\tau^2)]^{-1}$.
This sets a microscopic scale on which the averaged number of eigenvalues
in any domain of unit area remains finite
when $N \to \infty $. Remarkably, the $\tau$-dependence is
essentially trivial on this scale: the statistical properties
of eigenvalues are described by $\tilde R_n (z_1, \ldots , z_n)
\equiv N^{-n} R_n (\sqrt{N}Z_1, \ldots , \sqrt{N}Z_n)$ and
\begin{eqnarray}\label{Gin}
 \lim_{N\to \infty} \tilde R_n (z_1, \ldots , z_n)=
 \left[
\frac{1}{ \pi  (1-\tau^2)   }
\right]^n
e^{-
    \frac{1}{1-\tau^2}\sum_{j=1}^n |z_j|^2
   }
     \det
       \left[
       e^{
                   \frac{1}{1-\tau^2}z_jz_k^*
         }
     \right]_{j,k=1}^n.
\end{eqnarray}
This limiting relation can be inferred \cite{FKS3} from Mehler's formula
for the Hermite polynomials \cite{S}.
After the trivial additional rescaling  $z \to z\sqrt{1-\tau^2} $
the expression on the right-hand side in  Eq.\ (\ref{Gin})
becomes identical to that found by Ginibre \cite{Gin}.

Now we move on to the regime of weak non-Hermiticity.
We know that in this regime
new non-trivial correlations occur
on the scale:
$\mbox{Im} Z_{1,2}=O(1/N)$, $\mbox{Re} Z_1-\mbox{Re}
 Z_2=O(1/N)$.
Correspondingly, we
introduce new variables $x,y_1,y_2,\omega$ in such a way that:
$x=\mbox{Re}\left(Z_1+Z_2)\right/\! 2$,
$y_{1,2}=N\mbox{Im}\left(Z_{1,2} \right)$,
$\omega=N\mbox{Re}\left(Z_1-Z_2\right)$,
 and consider them finite when performing the limit $N\to
\infty$.

Substituting Eq.(\ref{H}) into Eq.(\ref{K}) and
using the above definitions
we can explicitly perform the limit $N\to \infty$,
taking into account that $\lim_{N\to \infty} N(1-\tau)= \alpha^2/2$.
The details of the procedure are given elsewhere \cite{FKS3}.
In this regime
\begin{eqnarray}\label{kern}
\lefteqn{
\lim_{N\to \infty}\frac{1}{N^2}
K_N\left(x+\frac{\omega/2+iy_1}{N}, x-\frac{\omega/2+iy_2}{N}\right)
        }\\  \nonumber
 & & =\frac{1}{\pi\alpha}
\exp{\left\{- \frac{y_1^2+y_2^2}{\alpha^2}+\frac{ix (y_1-y_2)}{2}
\right\}}
\int_{-\pi\nu_{sc}(x)}^{\pi\nu_{sc}(x)}\frac{du}{\sqrt{2\pi}}
\exp{\left[-\frac{\alpha^2u^2}{2}-u(y_1+y_2)+2i\omega u\right]},
\end{eqnarray}
with
$\nu_{sc}(X)=\frac{1}{2\pi}\sqrt{4-X^2}$ standing for the Wigner
semicircular density of
real eigenvalues of the Hermitian part $\hat{H}$ of the matrices $\hat{J}$.

Equation (\ref{kern})
 constitutes the most important result of the present section.
The kernel $K_N$ given by Eq.\ (\ref{kern})
determines all the properties of complex eigenvalues in the regime of
weak non-Hermiticity. For instance,
the mean value of the density
$\rho(Z)= \sum_{i=1}^N\delta^{(2)}(Z-Z_i)$
of complex eigenvalues $Z=X+iY$ is  given by $
\langle\rho(Z)\rangle= K_N(Z,Z^*)$.
Putting $y_1=y_2$ and $\omega=0$ in Eqs (\ref{kern})
we immediately recover the density
 Eq.(\ref{13}) found by the supersymmetry approach
\footnote{In the present section
we normalized $\hat{H}_2$ in such a way that for weak non-Hermiticity regime
we have
$\lim_{N\to\infty} \mbox{Tr}\hat{H}_2^2=N$, whereas
the nomalization Eq.(\ref{norm}) gives
$\lim_{N\to\infty} \mbox{Tr}\hat{H}_2^2=N(1+w^2)$.
It is just because of this difference the parameter
$\tilde{a}$ entering Eq.(\ref{13})
contains an extra factor $1+w^2$ as compared to the present case.}.

One of the most informative statistical measures of the spectral correlations
is the `connected' part of the two-point correlation function of eigenvalue  
densities:
\begin{eqnarray}\label{rr}
\left\langle \rho(Z_1)\rho(Z_2)\right\rangle_c
=\left\langle
\rho(Z_1)\right\rangle \delta^{(2)}(Z_1-Z_2)-{\cal Y}_2(Z_1,Z_2),
\end{eqnarray}
In particular, it determines the variance $\Sigma^2(D)=
\langle n(D)^2\rangle-\langle n(D)\rangle^2$ of the number
$n=\int_D d^2Z\rho(Z)$
of complex eigenvalues in any domain $D$ in the complex plane as:
\begin{eqnarray}
\Sigma_2(D)= \int_Dd^2Z_1\int_Dd^2Z_2[\langle\rho(Z_1)\rho(Z_2)\rangle-
\langle\rho(Z_1)\rangle\langle\rho(Z_2)\rangle]=
\int_Dd^2Z\langle\rho(Z)\rangle-\int_Dd^2Z_1\int_Dd^2Z_2
{\cal Y}_2(Z_1,Z_2)
\end{eqnarray}

Comparing Eq.(\ref{rr}) with the definition Eqs.\ (\ref{p_n})--(\ref{K})
we see that
the {\it cluster function} ${\cal Y}_2(Z_1,Z_2)$
is expressed in terms of the kernel $K_N$
as ${\cal Y}_2(Z_1,Z_2)=\left|K_N(Z_1,Z^*_2)\right|^2$.

It is evident that in the limit of weak non-Hermiticity
the kernel $K_N$
depends on $X$ only via the semicircular density $\nu_{sc}(X)$.
Thus, it does not change with $X$ on the local scale comparable with
the mean spacing along the real axis $\Delta\sim 1/N$.

The cluster function is given by the following explicit expression:
\begin{eqnarray}\label{clexp}
{\cal Y}(\omega,y_1,y_2)= \frac{N^4}{\pi^2 \alpha^2}
e^{-2\frac{y_1^2+y_2^2}{\alpha^2}}
\left|\int_{-\pi\nu(X)}^{\pi\nu(X)}\frac{du}{(2\pi)^{1/2}}
\exp{\left[-\frac{\alpha^2 u^2}{2}-u(y_1+y_2)+iu\omega\right]}\right|^2
\end{eqnarray}

The parameter
$a=\pi\nu(X)\alpha$ controls the deviation from Hermiticity
\footnote{ In our earlier Letter \cite{FKS2} we used the definition of the
parameter $a$ different by a factor of 2 from the present one.}.
When $a\to 0$ the cluster function tends to GUE form
${\cal Y}_2(\omega,y_1,y_2)=\frac{N^4}{\pi^2}\delta(y_1)
\delta(y_2)\frac{\sin^2{\pi\nu(X)\omega}}{\omega^2}$.
In the opposite case
$a\gg 1$ the limits of integration in Eq.(\ref{clexp})
can be effectively put to $\pm\infty$ due to the Gaussian cutoff
of the integrand. The corresponding Gaussian
integration is trivially performed
yielding in the original variables $Z_1,Z_2$
the expression equivalent (up to a trivial rescaling) to
that found by Ginibre \cite{Gin}:
${\cal Y}_2(Z_1,Z_2)=(N^2/\pi \alpha^2)^{2}
\exp\{-N^2|Z_1-Z_2|^2/{\alpha}^2\}$.

The operation of calculating the Fourier transform of the cluster function
over its arguments $\omega,y_1,y_2$ amounts to
 simple Gaussian and exponential integrations. Performing them
one finds
the following expression for the {\it spectral form-factor}:
\begin{eqnarray}\label{for}
b(q_1,q_2,k)=\int_{-\infty}^{\infty}d\omega
\int_{-\infty}^{\infty}dy_1\int_{-\infty}^{\infty}dy_2
{\cal Y}_2(\omega,y_1,y_2)\exp\{2\pi i(\omega k+y_1q_1+y_2q_2)\}\\ \nonumber
=N^4
\exp\{-\frac{\alpha^2}{2}\left(q_1^2+q_2^2+2k^2\right)\}
\frac{\sin{\left[\pi^2 \alpha^2(q_1+q_2)(\nu(X)-|k|)\right]}}
{\pi^2 \alpha^2(q_1+q_2)}\theta(\nu(X)-|k|)\,
\end{eqnarray}
where $\theta(u)=1$  for $u>0$ and zero otherwise.

We see, that everywhere in the regime of weak non-Hermiticity
$0<\alpha<\infty$ the formfactor shows a kink-like behavior
at $|k|=\nu(X)$. This feature is inherited from the corresponding
Hermitian counterpart-the Gaussian Unitary Ensemble. It
reflects the oscillations of the cluster function with
$\omega$ which is a manifestation of the long-ranged
order in eigenvalue positions along the real axis \cite{Bohigas}.
When non-Hermiticity increases
the oscillations become more and more damped.

 As we already discussed above the knowledge of the formfactor
 allows one to determine the variance $\Sigma_2$ of a number of
eigenvalues in any domain $D$ of the complex plane.
Small $\Sigma_2$ is a signature of a tendency for levels to form
a cristal-like structure with long correlations.
In contrast, increase in the number variance signals about growing
decorrelations of eigenvalues.

 In a general case
this expression is not very transparent, however. For this reason we
restrict ourselves to the simplest case,
choosing  the domain $D$ to be the infinite strip of width $L_x$
(in units of mean spacing along the real axis $\Delta=(\nu_{sc}(0)N)^{-1}$)
oriented perpendicular to the real axis:
$0<\mbox{Re}Z<~L_x\Delta;\quad -\infty
<\mbox{Im}Z<\infty$.
Such a choice means that we look only at real parts of complex
eigenvalues irrespective of their imaginary parts.
It is motivated, in particular, by
the reasons of comparison with the GUE case, for which  the function
$\Sigma(L_x)$ behaves at large $L_x$ logarithmically:
$\Sigma(L_x)\propto \ln{L_x}$ \cite{Bohigas}.

After simple calculations one finds
\footnote{In our earlier Letter \cite{FKS2} the expression
Eq.(\ref{var}) and formulae derived from it
 erroneously contained $\pi a$ instead of $a$.}

\begin{equation}
\label{var}
\Sigma_2(L_x)=
L_x\left[1-\frac{2}{\pi^2}\! \int_0^{L_x}\!
\frac{dk}{k^2}(1-\frac{k}{L_x})\sin^2{(\pi k)}
e^{-(\frac{a k}{L_x})^2 }\right]
\end{equation}

First of all, it is evident that $\Sigma_2$ grows systematically
with increase in the degree of non-Hermiticity $a=\pi\nu(0)\alpha$.
This fact signals on the gradual decorrelation of the {\it real}
parts Re$Z_i$ of complex eigenvalues. It can be easily understood
because of increasing possibility for eigenvalues to avoid one
another along the $Y={\mbox Im}Z$ direction, making their projections on the
real axis $X$ to be more independent.

In order to study the difference from
the Hermitian case in more detail let us consider
again the large $L_x$ behavior. In that case the upper limit of the integral
in Eq.(\ref{var}) can be set to infinity.
Then it is evident, that
 the number variance is only slightly modified by non-Hermiticity
as long as $a\ll L_x$. We therefore consider the case
$a\gg 1$ when we expect essential differences from the Hermitian
case.

 In a large domain $1\ll L_x\sim a$ the second term in the integrand of  
Eq.(\ref{var}) can be neglected and
the number variance grows like
 $\Sigma(L_x)=L_xf(L_x/a)$. We find it more transparent to rewrite
the function $f(u)$ in an equivalent form:
$$
f(u)=1+\frac{2}{\sqrt{\pi}}\left\{\frac{1}{2\pi u}\left(1-e^{-\pi^2u^2}\right)-
\int_{0}^{\pi u}dte^{-t^2}\right\}.
$$
which can be obtained from Eq.(\ref{var}) after a simple
transformation.

For $u=L_x/a\ll 1$ we have simply $f\approx 1$
 and hence a linear growth of the number variance. For $u\gg 1$ we have
$f~\approx~(\pi^{3/2}u)^{-1}$. Thus, $\Sigma_2(L_x)$ slows down:
$\Sigma_2(L_x)\approx \frac{a}{(\pi)^{3/2}}$.

Only for exponentially large $L_x$ such that
$\ln{(L_x/a)}\sim a$ second term in Eq.(\ref{var}) contributes significantly.
Calculating its contribution explicitly and
 remembering
that $\Sigma_2^{(1)}\mid_{L_x>>a}\approx a/(\pi^{3/2})$ we
finally find:
$$
\Sigma_2(L_x\gg a)= \frac{a}{\pi^{3/2}}+\frac{1}{\pi^2}
\left(\ln{\left(\frac{L_x}{ a}\right)}-\frac{\gamma}{2}\right)
$$
where $\gamma$ is Euler's constant. This logarithmic growth
of the number variance is reminiscent of that
typical for real eigenvalues of the Hermitian matrices.

Another important spectral characteristics which
can be simply expressed in terms of the cluster function is
the small-distance
 behavior of the nearest neighbor distance distribution
\cite{Mehta,Bohigas,Oas}.

We define the quantity $p(Z_0,S)$ as
 the probability density of the following
event: i) There is exactly one eigenvalue at the point $Z=Z_0$
of the complex plane. ii) Simultaneosly, there is exactly one
eigenvalue on the circumference of the circle $|Z-Z_0|=S$
iii) All other eigenvalues $Z_i$ are {\it out} of that circle:
$|Z_i-Z_0|>S$.

As a consequence, the normalization condition is: $\int d^2Z_0\int_0^{\infty}
dS\, p(Z_0,S)=1$. In particular, for Hermitian matrices with real
eigenvalues one has the relation: $p(Z_0,S)=\delta(\mbox{Im} Z_0)
\tilde{p}_X(S)$, with $\tilde{p}_X(S)$ being the conventional
"nearest neighbor spacing" distribution at the
point $X$ of the real axis \cite{Mehta}.

We are interested in finding the leading small-$S$ behavior
for the function $p(Z_0,S)$. It turns out to be given by
the following expression \cite{FKS3}:
\begin{equation}\label{small}
p(Z_0,S)\approx \frac{S}{N}\int_0^{2\pi}d\theta
\left[\langle\rho(Z_0)\rangle\langle\rho
\left(Z_0+Se^{i\theta}\right)\rangle-
{\cal Y}_2\left(Z_0,Z_0+Se^{i\theta}\right)\right]\end{equation}
where we used the definition of the cluster function, Eq.(\ref{rr}).

In the regime of weak non-Hermiticity this
formula is valid as long as the parameter $S$ is small
in comparison with a typical separation between real eigenvalues
of the Hermitian counterpart: $S\ll \Delta\sim 1/N$.

Substituting the expression Eqs.(\ref{13},\ref{clexp}) for the mean
density and the cluster function into Eq.(\ref{small})
 one arrives after
a simple algebra to the probability density to have one
eigenvalue at the point $Z_0=X+iy_0\Delta$ and its
closest neighbor at the distance $|z_1-z_0|=s\Delta,\,\,
\Delta=(\nu(X)N)^{-1}$, such that
$s\ll 1$:
\begin{eqnarray} \label{nns}
\nonumber
p_{\alpha}(X+iy_0\Delta,s\Delta)|_{s\ll 1}=
\frac{\pi\nu^2(X)}{2}\left[
g_{a}(y_0)\frac{\partial^2}{\partial y_0^2}g_{a}(y_0)-\left(
\frac{\partial}{\partial y_0}g_{a}(y_0)\right)^2\right]
 \displaystyle{e^{-4\frac{y_0^2}{a^2}}}\frac{s^3}{a^2}
\int_0^{\pi}d\theta
\exp{\left[-\frac{2}{a^2}(s^2\cos^2{\theta}-2y_0s\cos{\theta})\right]}
\end{eqnarray}
where
$$
g_a(y)=\int_{-1}^1\frac{du}{(2\pi)^{1/2}}\exp\{-\frac{a^2u^2}{2}-2uy\}
$$

First of all it is easy to see that in the limit $a\gg 1$
one has: $p_{a\gg 1}(Z_0,s\ll 1)
=\frac{2}{\pi} (s/a^2)^3$ in agreement with the cubic repulsion generic
for strongly non-Hermitian random matrices \cite{Gin,diss,Oas}.
On the other hand one can satisfy oneself that in
 the limit $a\to 0 $ we are back to
the familiar GUE quadratic level repulsion: $p_{a\to 0}(Z_0,s\ll 1)
\propto \delta(y_0) s^2$.
In general, the expression Eq.(\ref{nns})
describes a smooth crossover between the two regimes, although
for any $a\ne 0$ the repulsion is always cubic for $s\to 0$.

To this end, an interesting situation may occur when
 deviations from the Hermiticity are very weak: $a\ll \sqrt{2}$ and
`observation points' $Z_0$ are situated sufficiently far
from the real axis: $2|y_0|/a\gg 2^{-1/2}$.

Under this condition the following three regions for the
parameter $s$ should be distinguished:
i)$\frac{s}{a}\ll \frac{a}{4|y_0|}$
ii)$\frac{a}{4|y_0|}\ll \frac{s}{a}\ll 2\frac{|y_0|}{a}$
and finally iii) $ 2^{-1/2}\ll 2\frac{|y_0|}{a}\ll \frac{s}{a}\ll a^{-1}$.

In the regimes i) and ii) the term linear in $\cos{\theta}$
in the exponent of Eq.(\ref{nns}) dominates yielding the
result of integration to be the modified Bessel function
$\pi I_0\left(\frac{4y_0s}{a^2}\right)$. In the regime iii)
the term quadratic in $\cos{\theta}$ dominates producing
$2\pi e^{-(s/a)^2}I_0\left[(s/a)^2\right]\approx
\left(2\pi a/s\right)^{1/2}$.
As the result, the distribution
$p(Z_0,s)$ displays the following behavior:
\begin{eqnarray}\label{5/2}
\nonumber
p_{\alpha}(Z_0,s)=\frac{\pi^2\nu^2}{2}
\left[
g_{0}(y_0)\frac{\partial^2}{\partial y_0^2}g_{0}(y_0)-\left(
\frac{\partial}{\partial y_0}g_{0}(y_0)\right)^2\right]
e^{-4\frac{y_0^2}{a^2}}\left\{\begin{array}{cc}
\frac{s^3}{a^2}& \mbox{ for} \quad \frac{s}{a}\ll \frac{a}{4|y_0|}\\
 \frac{s^{5/2}}{2a\sqrt{2\pi|y_0|}} &\mbox{ for} \quad \frac{a}{4|y_0|}\ll
\frac{s}{a}\ll 2\frac{|y_0|}{a}\\ \sqrt{\frac{2}{\pi}}
\frac{s^2}{a} &\mbox{ for} \quad 2\frac{|y_0|}{a}\ll \frac{s}{a}\ll a^{-1}
\end{array}\right.,\end{eqnarray}
with $g_0(y)\equiv g_a(y)|_{a=0}$.

Unfortunately, it might be a very difficult task to detect numerically
the unusual power law $p(s)\propto s^{5/2}$
 because of
the low density of complex eigenvalues in
the observation points reflected by the presence of the
Gaussian factor in the expression Eq.(\ref{5/2}).

\section{Conclusion}

In the present paper we addressed the issue of eigenvalue statistics
of large weakly non-Hermitian matrices.

Our original motivation came from the field of resonance chaotic scattering.
The resonances, which are complex poles of the scattering matrix
enter the theory as complex eigenvalues of a non-Hermitian effective  
Hamiltonian of a particular type: ${\cal H}_{ef}=\hat{H}-i\hat{\Gamma}$.
We demonstrated that one can extract mean density of such
poles employing a mapping onto the supermatrix non-linear $\sigma-$model.
We also have shown how the resolvent of the non-Hermitian Hamiltonian ${\cal  
H}_{ef}$ can be used to describe the process of chaotic photodissociation  
and
presented the crossection autocorrelation function.

Guided by our experience with the resonances, we found
a regime of weak
non-Hermiticity for other types of non-selfadjoint random matrices.
The regime can be defined as that for which the imaginary
part $\mbox{Im} Z$ of a typical complex eigenvalue is of the same order
as the mean eigenvalue separation $\Delta$ for the corresponding Hermitian
counterpart.

Exploiting a mapping to the non-linear $\sigma-$model we are able
to show that there are three different "pure" symmetry classes of weakly
non-Hermitian matrices: i) almost Hermitian with complex entries
ii) almost symmetric with real entries and iii) complex symmetric
ones. Within each of these classes the eigenvalue statistics
is {\it universal} in a sense that it is the same irrespective of the
particular distribution of matrix entries up to
an appropriate rescaling. There are also crossover regimes between
all three classes.

Our demonstration of universality was done explicitly for
the density of complex eigenvalues of matrices
with independent entries. Within the non-linear $\sigma-$model formalism
one can easily provide a heuristic proof of such a universality
for higher correlation functions as well as for "rotationally
invariant" matrix ensembles, see \cite{HW}.
The above feature is a great advantage of the supersymmetry technique.

A weak point of that method is a very complicated representation of
the ensuing quantities. It seems, that the explicit evaluation of the
higher correlation functions is beyond our reach at the moment, and
even a calculation of the mean density requires a lot of effort, see
\cite{FKS1,Efnonh}. As a result, at present time
the mean density is known explicitly only
for the cases i) and ii).

Fortunately, because of the mentioned universality
another strategy can be pursued. Namely, one can concentrate
on the particular case of matrices with independent, Gaussian
distributed entries for which alternative analytical techniques might
be available. Such a strategy turned out to be a success for
the simplest case of complex almost-Hermitian matrices, where
we found the problem to be an exactly soluble one by the method of
orthogonal polynomials. This fact allowed us to extract all the
correlation functions in a mathematically rigorous way \cite{FKS2,FKS3}.

One might hope that combining the supersymmetric method
and the method of orthogonal polynomials one
will be able to elevate our understanding
of properties of almost-Hermitian
random matrices to the level typical for their Hermitian counterparts.

From this point of view a detailed numerical investigation of
different types of almost-Hermitian random matrices is highly
desirable. Recently, an interesting work in this direction appeared
motivated again by the theory of chaotic scattering \cite{reso}.
Unfortunately, matrices ${\cal H}_{ef}$ emerging in that theory are
different from the Gaussian matrices because of the
specific form of the antihermitean perturbation $i\hat{\Gamma}$ necessary to  
ensure the unitarity of the scattering matrix.
This fact makes impossible a quantitative comparison of our results with those 
obtained in \cite{reso}. The qualitative fact of increase in number
variance with increase in non-Hermiticity agrees well with our findings.

The author is much obliged to H.-J. Sommers,
B.A.Khoruzhenko and Y.Alhassid for
collaboration on different aspects of resonance scattering and
non-Hermitian random matrices considered in the present paper and
to D.Savin, J.Main, J.J.M.Verbaarschot and especially to V. Sokolov
for useful discussions.

The author is grateful to the organizers of the program "Disordered Systems  
and Quantum Chaos"
for the financial support of his stay at the Newton Institute, Cambridge.

The work was supported by SFB 237 "Unordnung und grosse Fluktuationen"
and EPRSC Research Grant GR/L31913.
The warm hospitality of the School of Mathematical Sciences, Queen Mary\&
Westfield
College, University of London, of the Sloane Physics Laboratory,
 Yale University
where different parts of the work were done is acknowledged with
thanks.


\begin{thebibliography}{99}

\bibitem{AlSi} B.L.Altshuler and B.D.Simon in: {\it Mesoscopic
Quantum Physics} ed. by E.Akkermans et al, Les Houches Summer School,  
Session LXI 1994, edited E.Akkermans et  al., Elsever Science.
\bibitem{Bohigas} O.Bohigas, in {\it Chaos and Quantum Physics. Proceedings of
the  Les-Houches Summer School. Session LII}, ed. by M.J.Giannoni et.al (North
Holland, Amsterdam, 1991), p.91
\bibitem{per}  M.Berry, {\it Proc.R.Soc.London}, Ser.A {\bf 400}, 229 (1985);
\bibitem{bogomol}  E.Bogomolny and J.Keating, {\it Phys.Rev.Lett} {\bf 77},
1472 (1996)
\bibitem{MK}  B.A.Muzykantsky and D.E.Khmelnisky {\it JETP Lett.} {\bf 62}
, 76 (1995);
\bibitem{aaas}  A.Andreev, O. Agam, B. Altshuler and B.Simons {
Phys.Rev.Lett.}, {\bf 76}, 3947 (1996)
\bibitem{Guhr} T.Guhr, A.M\"{u}ller-Groelling and
H.A.Weidenm\"{u}ller, to appear in Rev.Mod.Phys.
\bibitem{Sok} V.V.Sokolov and V.G.Zelevinsky  Phys.Lett.B {\bf 202},
10 (1988); Nucl.Phys.A {\bf 504}, 562 (1989);
F.Haake, F.Izrailev, N.Lehmann, D.Saher, and H.-J.Sommers,
Z.Phys.B {\bf 88}, 359 (1992);  N.Lehmann, D.Saher,
V.V.Sokolov, and H.-J.Sommers,
Nucl.Phys.A {\bf 582}, 223 (1995); M. M\"{u}ller, F.-M.Dittes, W.Iskra,
and I. Rotter, Phys.Rev.E {\bf 52}, 5961 (1995).
\bibitem{FS} Fyodorov Y V and Sommers H-J  {\it Pis'ma ZhETF} v.63,970 (1996);
[{\it JETP Letters} v.63 ,1026 (1996)]
\bibitem{reso} T.Gorin, F.-M.Dittes, M.M\"{u}ller, I.Rotter and
T.H.Seligman , {\it Phys.Rev.E}, v.56 , 2481 (1997)
\bibitem{HN} G.Hackenbroich and J.N\"{o}ckel {\it Europh.Lett.} v.39, 371
(1997)
\bibitem{Main} J.Main and G.Wunner {\it J.Phys.B: At.Mol.},
{\bf 27}, 1994 (1994); B.Gremaud and D.Delande {\it Europh.Lett.} v.40 ,  
p.363 (1997)
\bibitem{Blumel} R.Blumel, {\it Phys.Rev.E}, {\bf 54}, 5420 (1996);
\bibitem{Man} V.A.Mandelshtam and H.S.Taylor, {\it J.Chem.Soc.Faraday.Trans.
}, {\bf 93} ,847 (1997) and {\it Phys.Rev.Lett.}, {\bf 78}, 3274
(1997)
\bibitem{Smilansky} U.Smilansky
 in {\it Chaos and Quantum Physics. Proceedings of
the  Les-Houches Summer School. Session LII}, ed. by M.J.Giannoni et.al (North
Holland, Amsterdam, 1991),p.372
\bibitem{FSR}  Y.V.Fyodorov and H.-J.Sommers , {\it J.Math. Phys.}, {\bf 38}
, 1918 (1997)
\bibitem{VWZ} J.J.M.Verbaarschot, H.A.Weidenm\"{u}ller, M.R.Zirnbauer
, {\it Phys.Rep.} v.129 , 367 (1985)
\bibitem{Mah} C.Mahaux and H.A.Weidenm\"{u}ller, {\it Shell Model Approach
in Nuclear Reactions} (North Holland, Amsterdam), 1969
\bibitem{Kob} I.Yu.Kobzarev, N.N.Nikolaev and L.B.Okun, Yad.Phys. {\bf 10}, 864
(1969) [in Russian]
\bibitem{Liv} M.S.Livsic {\it Operators, Oscillations, Waves: Open
Systems}, \\
Amer.Math.Soc.Trans. v.34 (Am.Math.Soc.,Providence,RI,1973)
\bibitem{Efbook} K.B. Efetov {\it Supersymmetry in Disorder and Chaos}
(Cambridge University Press,1996 )
\bibitem{my} Y.V. Fyodorov  in ``Mesosocopic Quantum Physics'',
 Les Houches Summer School, Session LXI,1994, edited E.Akkermans et
al., Elsever Science, p.493
\bibitem{FA} Y.V.Fyodorov and Y.Alhassid, under preparation
\bibitem{Sav} D.V.Savin and V.V.Sokolov, {\it Phys.Rev.E} v.56, R4911
(1997)
\bibitem{ewa} E.Gudowska-Nowak, G.Papp and J.Brickmann {\it
Chem.Physics} v.220, 125 (1997)
\bibitem{diss} R.Grobe, F.Haake,  and H.-J.Sommers,
Phys.Rev.Lett. {\bf 61}, 1899 (1988)
\bibitem{Haake_book} F. Haake
{\it Quantum Signature of Chaos} (Berlin, Springer,  1991 )
\bibitem{reichl} L.E.Reichl, Z.Y.Chen and M.Millonas {\it
Phys.Rev.Lett.} v.63, 2013 (1989)
\bibitem{neural}  H.-J.Sommers, A.Crisanti, H.Sompolinsky,
 and Y.Stein, Phys.Rev.Lett. {\bf 60}, 1895 (1988);
H. Sompolinsky , A. Crisanti  and H.-J. Sommers
{\it Phys. Rev. Lett.} {\it 61}  259, 1988;
B. Doyon ,B. Cessac , M. Quoy  and M. Samuelidis
 {\it Int. J. Bifurc. Chaos} {\bf 3} 279 (1993);
\bibitem{Nels} N.Hatano~and~D.R.Nelson,~Phys.Rev.Lett.~{\bf 77},~570
(1996); {\it Phys.Rev.B} v.56 (1997), 8651
\bibitem{nhloc} P.W.Brouwer, P.G.Silvestrov and C.W.J. Beenakker
{\it Phys.Rev.B} v.56 (1997), 4333;  R.A.Janik et al.,
e-preprint cond-mat/9705098; B.A. Khoruzhenko and I. Goldscheid,
e-preprint cond-mat/9707230;
\bibitem{Efnonh}  K.B.Efetov, {\it Phys.Rev.Lett.} {\bf 79}, 491 (1997)
\bibitem{pass} J. Miller and J. Wang, {\it Phys. Rev. Lett.} {\bf 76}, 1461
(1996); J. Chalker and J. Wang,{\it Phys.Rev.Lett.} {\bf 79} , 1797 (1997)
\bibitem{QCD} M.A. Stephanov,  {\it Phys. Rev. Lett.} {\bf 76}, 4472 (1996);
R.A.Janik et al., {\it Phys. Rev. Lett.} {\bf 77}, 4876
(1996); M.A. Halasz, A.D. Jackson and J.J.M. Verbaarschot , {\it
Phys.Rev.D}, v.56 , 5140 (1997);
M.A. Halasz, J.C.Osborn and J.J.M. Verbaarschot e-preprint hep-th/9704007.
\bibitem{int} T. Akuzawa and M. Wadati, {\it J. Phys. Soc. Jpn.} {\bf 65 },
1583 (1996).
\bibitem{cal} M.V. Feigelman and M.A Skvortsov, {\it Nucl. Phys.B} {\bf 506}  
(1997), 665;
A. Khare and K. Ray, e-preprint hep-th/9609025.
\bibitem{Khor}B.A.Khoruzhenko, J.Phys.A {\bf 29}, L165 (1996).
\bibitem{FKS1} Y.V. Fyodorov, B. Khoruzhenko and H.-J. Sommers, {\it
Physics Letters A} {\bf 226}, 46 (1997);
\bibitem{FKS2}Y.V. Fyodorov, B. Khoruzhenko and H.-J. Sommers,
{\it Phys.Rev.Lett.} v. 79, 557 (1997)
\bibitem{Oas} G. Oas, {\it Phys. Rev. E} {\bf 55}, 205 (1997)
\bibitem{free} R.A. Janik, M.Nowak,G.Papp and I.Zahed
Nucl.Phys.B {\bf 501}, 603 (1997); J. Feinberg and  A. Zee,
Nucl.Phys.B. {\bf 501},643(1997) and Nucl.Phys.B. {\bf 504}, 579 (1997)
\bibitem{Kus} M. Kus, F. Haake, D. Zaitsev and A.Huckleberry,
{\it J.Phys.A } to be published.
\bibitem{Gin} J. Ginibre, {\it J. Math. Phys.} { \bf 6}, 440 (1965).
\bibitem{Mehta} M.L.Mehta, {\it Random Matrices} (Academic Press Inc., N.Y.,
1990)
\bibitem{Gir} V. Girko, {\it Theor. Prob. Appl.} {\bf 30}, 677 (1986).
\bibitem{Lehm1} N.Lehmann and H.-J.Sommers, Phys.Rev.Lett.
{ \bf 67}, 941 (1991).
\bibitem{Forr} P.J. Forrester, {\it Phys. Lett. A} { \bf 169},21 (1992);
{\it J. Phys. A:Math. Gen.} {\bf 26}, 1179 (1993).
\bibitem{Edel} A. Edelman, {\it J. Multivariate Anal.} {\bf 60}, 203
(1997); A.Edelman, E.Kostlan and M.Shub {\it J.Am.Math.Soc.} v.7
, 247 (1994)
\bibitem{Bai} Z. D. Bai, {\it Ann. Prob. } {\bf 25}, 494 (1997).
\bibitem{FKS3}Y.V. Fyodorov, B. Khoruzhenko and H.-J. Sommers, {\it
Annales de l'Institut Henri Poincare}, to be published
\bibitem{Lew} C.H.Lewenkopf and H.A.Weidenm\"{u}ller, {\it Ann.Phys.} v.212  
,53 (1991)
\bibitem{Gasp} P.Gaspard in "Quantum Chaos", {\it Proceedings of
E.Fermi Summer School,1991} ed. by G.Casati et al. (North Holland,
Amsterdam,1991), p.307
\bibitem{Stone} D.Stone in {\it Mesoscopic Quantum Physics}, see \cite{AlSi}
\bibitem{Schrev} R.Schinke, H.-M.Keller, M.Stumpf and A.J.Dobbyn
, {\it J.Phys.B: At.Mol.}, {\bf 28}, 2928 (1995)
\bibitem{Flam} V.V.Flambaum, A.A.Gribakina and G.F.Gribakin,
 {\it Phys.Rev.A} v.54, 2066,(1996)
\bibitem{Mol} P.A.Moldauer, {\it Phys.Rev.} v.157,907 (1967);
M.Simonius {\it Phys.Lett.} v.52B, 279 (1974)
\bibitem{Seba} S.Albeverio,F.Haake,P.Kurasov,M.Kus and P.Seba
, {\it J.Math.Phys.} v.37, 4888 (1996)
\bibitem{Harney} H.L.Harney, F.M.Dittes and A. M\"{u}ller
{\it Ann.Phys.} v.220, 159 (1992)
\bibitem{td1}  N.Lehmann , D.Savin, V.V. Sokolov and H.-J. Sommers
 {\it Physica D} {\bf 86} 572 (1995); Y.V. Fyodorov, D.Savin and
H.-J.Sommers , {\it Phys.Rev.E}, {\bf 55},4857 (1997)
\bibitem{Schinke} R.Schinke {\it Photodissociation Dynamics},
(Cambridge University, 1993);
\bibitem{Sok1} V.V.Sokolov and V.G.Zelevinsky {\it Phys.Rev.C} {\bf 56}
 ,311 (1997);
 \bibitem{Tan} N. Taniguchi,A.V. Andreev and B.L. Altshuler , {\it
Europh.Lett.} , {\bf 29}, 515 (1995)
 \bibitem{cross} A.Pandey and M.L.Mehta, {\it Commun.Math.Phys.} v.87
, 449 (1983)
\bibitem{Alt} A.Altland, K.B.Efetov, S.Iida, {\it
J.Phys.A:Math.Gen} v.26 ,2545 (1993)
\bibitem{MF} A.D.Mirlin, Y.V.Fyodorov, {\it J.Phys.A} v.24 , 2273 (1991);
Y.V.Fyodorov, A.D.Mirlin, {\it Phys.Rev.Lett.} v.67 , 2049 (1991)
\bibitem{FSC}Y.V.Fyodorov and  H.-J.Sommers,{\it Z.Phys.B} v.99 ,123 (1995)


\bibitem{note}
Strictly speaking, the form of the correlation function
of eigenvalue densities for sparse matrices
was shown to be identical to that known for the corresponding Gaussian
ensemble provided $p$ exceeds some critical value $p=p_{l}$. The
"threshold" value $p_{l}$ is nonuniversal and
depends on the form of the distribution ${\cal P}(\hat{H})$ \cite{MF}.
However, direct numerical simulations, see S.Evangelou {\it J.Stat.Phys.}  
v.69 (1992), 361 show that actual value is $1<p_l<2$.
Thus, even existence of two nonvanishing elements
per row already ensure, that the corresponding
statistics belongs to the Gaussian universality class.
In the present paper we assume that $p>p_l$.
\bibitem{HW} G. Hackenbroich and H.A. Weidenm\"{u}ller
{\it Phys.Rev.Lett.} {\bf 74} 4118 (1995)
\bibitem{orth} F. Di Francesco, M.Gaudin, C.Itzykson, and F.Lesage
Int.J.Mod.Phys.A {\bf 9}, 4257 (1994).
\bibitem{FJ}  P.J.Forrester and B.Jancovici, Int.J.Mod.Phys.A {\bf 11},941 (1997)
\bibitem{S} G.Szeg\"o, {\it Orthogonal polynomials}, 4th ed. (AMS, Providence,
1975), p. 380.
\bibitem{GR} I.S.Gradshteyn, I.M.Ryzhik "Table of
Integrals, Series, and Products" (Academic Press,
N.Y. 1980).

\end{thebibliography}
\end{document}